\documentclass{eas}

\usepackage{graphicx}
\usepackage{hyperref}
\usepackage{astron}

\TitreGlobal{The Milky Way Unravelled by Gaia}

\begin{document}

\title{Gaia Astrometric Science Performance -- Post-Launch Predictions}
\runningtitle{Post-Launch Astrometric Science Performance}
\author{J.H.J. de Bruijne}\address{Scientific Support Office, Directorate of Science and Robotic Exploration, European Space Research and Technology Centre (ESA/ESTEC), Keplerlaan 1, 2201AZ, Noordwijk, The Netherlands\ \email{jos.de.bruijne@esa.int}}
\author{K.L.J. Rygl}\sameaddress{1}
\author{T. Antoja}\sameaddress{1}

\begin{abstract}
The standard errors of the end-of-mission Gaia astrometry have been re-assessed after conclusion of the in-orbit commissioning phase of the mission. An analytical relation is provided for the parallax standard error $\sigma_\varpi$ as function of Gaia $G$ magnitude (and $V-I$ colour) which supersedes the pre-launch relation provided in de Bruijne \cite*{2012Ap&SS.341...31D}.
\end{abstract}

\maketitle

\section{Introduction}

After the success of the Hipparcos astrometry mission \cite{Perryman2009}, culminating in the Hipparcos and Tycho Catalogues \cite{hip:catalogue}, ESA has recently launched the Gaia spacecraft \cite{2012AN....333..453P}. Gaia's main goal is to unravel the formation history, evolution, structure, and dynamics of our Galaxy, the Milky Way, through measuring the positions, motions, and astrophysical parameters of the brightest 1,000 million stars in the sky. In addition, Gaia will provide unique contributions to the study of solar-system bodies \cite{2007A&A...474.1015T,2012P&SS...73....5T}, fundamental physics \cite{2010IAUS..261..306M}, external galaxies \cite{2013A&A...556A.102K}, the reference frame \cite{2013A&A...552A..98T}, the transient universe \cite{2012arXiv1210.5007W,2014sf2a.conf..445T}, stellar structure, physics, and evolution \cite{2013A&A...559A..74B}, variable stars \cite{2014IAUS..298..265E}, double and multiple stars \cite{2011AIPC.1346..122P}, young, massive stars (Rygl et al., this volume), brown dwarfs \cite{2014arXiv1404.3896D}, white dwarfs \cite{2014arXiv1411.5206J}, open clusters \cite{2011sca..conf....0A}, globular clusters \cite{2013MmSAI..84...83P}, the distance scale \cite{2012Ap&SS.341...15T}, exo-planets \cite{2012ApJ...753L...1D,2014MNRAS.437..497S,2014ApJ...797...14P}, and so forth. Gaia will continuously scan the heavens during its five-year nominal lifetime, which started on 25 July 2014, and the first, intermediate catalogue is expected in 2016 (\url{http://www.cosmos.esa.int/web/gaia/release}). A review of the Gaia mission, including an overview of its payload, is given in de Bruijne et al.\ \cite*{2010SPIE.7731E..1CD}. The main Gaia website is \url{http://www.cosmos.esa.int/gaia}.

\begin{figure*}[t]
  \centering
  \includegraphics[trim=35 0 78 27,clip,width = 0.9\textwidth]{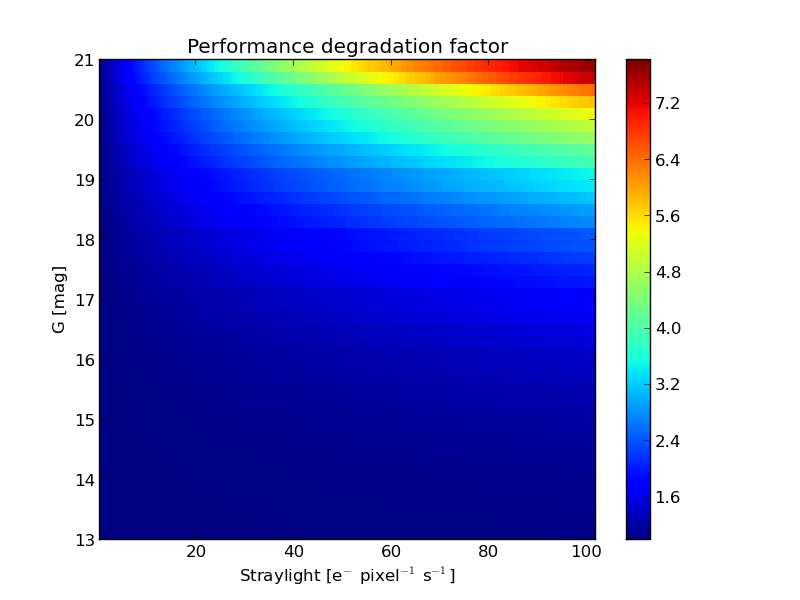} 
  \caption{End-of-mission parallax-standard-error degradation factor, colour coded on a linear scale, as function of straylight level and $G$ magnitude. For instance: with a hypothetical, continuous straylight level of 80~e$^-$~pixel$^{-1}$~s$^{-1}$ on all astrometric CCDs, Gaia's parallax standard error at $G=20$~mag would degrade by a factor 4 (turquoise colour). For low straylight levels and/or bright stars, the performance impact is small.}
  \label{fig:parametric_study}
\end{figure*}

\section{Commissioning}\label{sec:com}

The expected characteristics of the Gaia data have been subject to numerous studies \cite{2012A&A...543A.100R,2014A&A...566A.119L}. The particular question of how the expected standard errors of single, well-behaved stars depend on magnitude and colour of the source -- as well as on position in the sky -- has been answered by de Bruijne \cite*{2012Ap&SS.341...31D}; the expected astrometric correlations have been discussed in detail by Holl et al.\ \cite*{2012A&A...543A..14H,2012A&A...543A..15H}. These studies, however, were concluded prior to Gaia's launch and are hence based on the pre-launch predictions of the telescope and payload behaviour. During the in-orbit commissioning phase of the mission, however, three surprises showed up (Prusti, this volume):
\begin{enumerate}
\item there is significant stray light, which periodically varies with time;
\item the transmission of the optics slowly degrades with time (currently at a rate of $\sim$40~mmag per 100~days) as a result of contamination by water ice;
\item the intrinsic instability of the basic angle -- which separates the lines of sight of the two telescopes -- is larger than expected.
\end{enumerate}
Issue 2 is kept under control by (semi-)periodically heating the payload to decontaminate the optics and only impacts the end-of-life performance at a level of $\sim$$10$\%; this can hence be considered to be covered by the $20$\% ``science margin'' that is included in all performance predictions. Issue 3 is kept under control through the Basic-Angle-Monitor device \cite{2014SPIE.9143E..0XM} which allows to measure the variations in the basic angle and inject this information into the astrometric global iterative solution \cite{2012A&A...538A..78L}. Issue 1, on the other hand, is accompanied with increased noise levels and hence {\em does} lead to an irreversible degradation of the end-of-life astrometric (and also photometric and spectroscopic -- Cropper et al., this volume) standard errors, as quantified in this paper.

\section{Post-launch estimates}

\begin{figure*}[t]
  \centering
  \includegraphics[trim=0 0 0 0,clip,width = 0.75\textwidth]{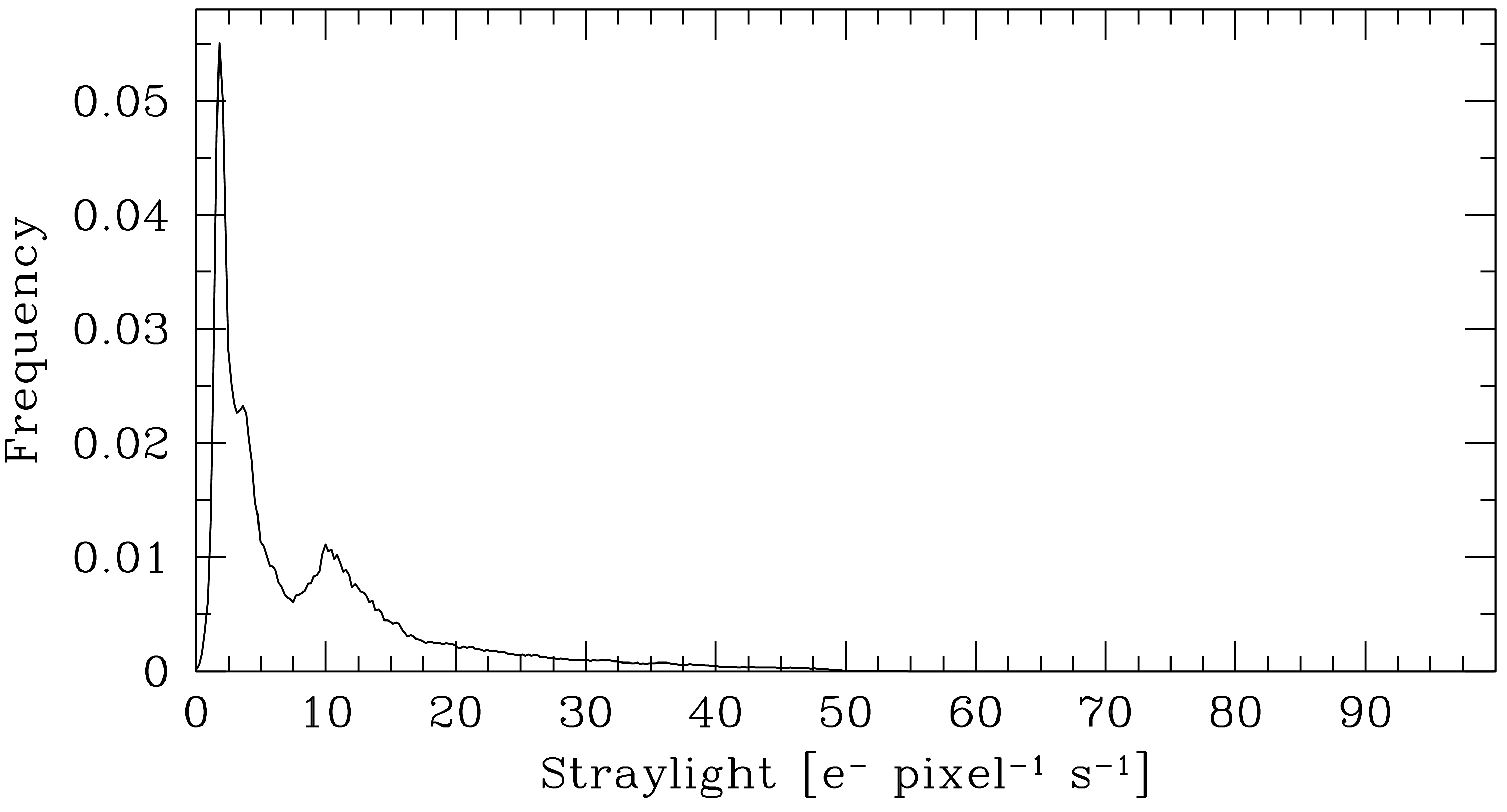} 
  \caption{Frequency distribution of measured straylight levels during the in-orbit commissioning phase. Measured values in excess of 100~e$^-$~pixel$^{-1}$~s$^{-1}$ have been added to the final bin (not apparent in the figure). The median straylight level is 5.5~e$^-$~pixel$^{-1}$~s$^{-1}$.}
  \label{fig:observed_statistics}
\end{figure*}

The Gaia Accuracy Analysis Tool has been run for 451 straylight levels ranging between 0 (no straylight) and 100~e$^-$~pixel$^{-1}$~s$^{-1}$. For each straylight level, 300 Monte-Carlo centroiding experiments have been performed for each combination of the: two telescopes, 62 astrometric CCD detectors, 16 stars with various spectral types ranging from B1V to M6V, two different interstellar extinctions (0 and 5 mag), and 9 different $V$ magnitudes between $13$ and $21$ mag.

The results of the $\sim$$4.8$~billion centroiding experiments have been re-arranged and averaged to provide the ratio of the CCD-level centroiding error at a certain straylight level compared to the reference case of no straylight, referred to as the ``performance-degradation factor'', as function of Gaia white-light $G$ magnitude as defined in Jordi et al. \cite*{2010A&A...523A..48J}. The performance-degradation factor as function of straylight level and $G$ magnitude is displayed in Figure~\ref{fig:parametric_study}. For instance, it follows that, with a hypothetical, omni-present straylight level of 80~e$^-$~pixel$^{-1}$~s$^{-1}$ during the full mission duration, Gaia's parallax standard error at $G = 20$~mag would degrade by a factor 4 (turquoise colour).

\begin{figure*}[t]
  \centering
  \includegraphics[trim=30 0 90 80,clip,width = \textwidth]{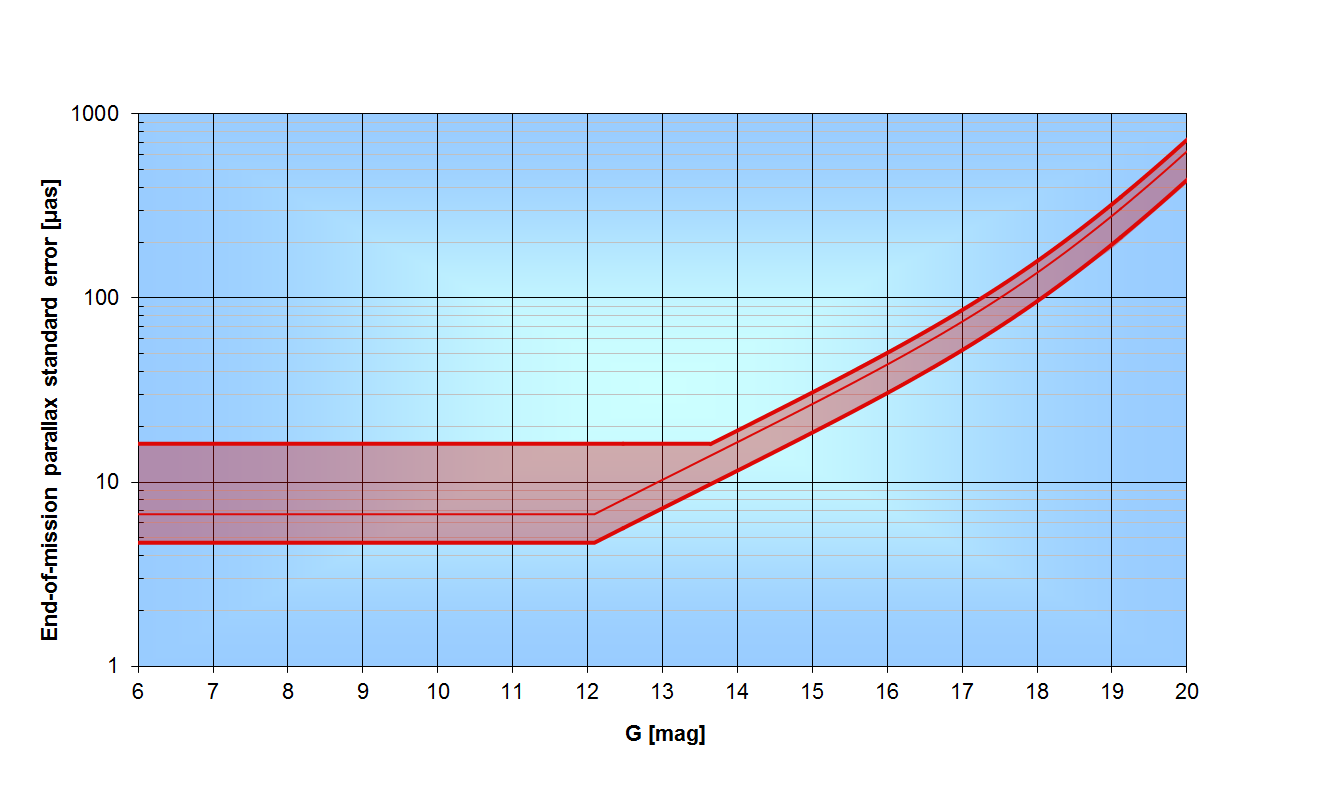} 
  \caption{Sky-average, end-of-mission, parallax standard error, in units of $\mu$as, as function of Gaia $G$ magnitude for an unreddened G2V star ($V-I=0.75$~mag and $V-G=0.16$~mag). The middle red curve refers to the sky-average dependence; the upper and lower curves -- which bound the shaded area -- denote the expected variations caused by position on the sky, star colour, and bright-star observing conditions (TDI gates, on-board magnitude-estimation errors, etc.). The slight upturn of the linear relation in log space starting around $\sim$17~mag is mainly caused by straylight.}
  \label{fig:curve}
\end{figure*}

\begin{table}[t]
  \caption{Sky-average, end-of-mission, astrometric standard errors -- $\sigma_0$ in $\mu$as for position at mid-epoch, $\sigma_\varpi$ in $\mu$as for parallax, and $\sigma_\mu$ in $\mu$as~yr$^{-1}$ for proper motion -- as function of Gaia $G$ magnitude for an unreddened G2V star ($V-I=0.75$~mag and $V-G=0.16$~mag). For stars in the range $G=3$--$12.09$~mag, the numbers refer to ``average errors'' (see text).}\label{tab:errors}
  \begin{center}
    \begin{tabular}[h]{cccccccccc}
      \hline\hline
      \\[-8pt]
      $G$ [mag] & $3$--$12.09$ & 13 & 14 & 15 & 16 & 17 & 18 & 19 & 20\\
      \\[-10pt]
      \hline
      \\[-8pt]
      $\sigma_0$ [$\mu$as] & 5.0 & 7.7 & 12.3 & 19.8 & 32.4 & 55.4 & 102 & 208 & 466\\
      \\[-10pt]
      $\sigma_\varpi$ [$\mu$as]& 6.7 & 10.3 & 16.5 & 26.6 & 43.6 & 74.5 & 137 & 280 & 627\\
      \\[-10pt]
      $\sigma_\mu$ [$\mu$as~yr$^{-1}$] & 3.5 & 5.4 & 8.7 & 14.0 & 22.9 & 39.2 & 72.3 & 147 &330\\
      \\[-10pt]
      \hline
    \end{tabular}
  \end{center}
\end{table}

The distribution of straylight levels measured during commissioning, in units of e$^-$~pixel$^{-1}$~s$^{-1}$ and averaged over all pixel columns of all CCDs in the astrometric focal plane, is displayed in Figure~\ref{fig:observed_statistics}. The median straylight level is 5.5~e$^-$~pixel$^{-1}$~s$^{-1}$ and only 10\% of the distribution has straylight levels in excess of 20~e$^-$~pixel$^{-1}$~s$^{-1}$. By combining the results displayed in Figures~\ref{fig:parametric_study} and \ref{fig:observed_statistics}, the predicted, sky-average, end-of-mission parallax standard error as function of $G$ magnitude as displayed in Table~\ref{tab:errors} and in Figure~\ref{fig:curve} follows.

\section{Parametrisation}

We find that, following the pre-launch investigation \cite{2012Ap&SS.341...31D}, the predicted, post-launch, sky-average, end-of-mission parallax standard error is well represented by (\url{http://www.cosmos.esa.int/web/gaia/science-performance}):
\begin{equation}
\sigma_\varpi [\mu{\rm as}] = (-1.631 + 680.766 \cdot z + 32.732 \cdot z^2)^{1/2} \cdot [0.986 + (1 - 0.986) \cdot (V-I)],
\end{equation}
where
\begin{equation}
z = {\rm MAX}[10^{0.4 \cdot (12.09 - 15)}, 10^{0.4 \cdot (G - 15)}].\label{eq:z}
\end{equation}
The $V-I$ colour term represents the widening of the point spread function at longer wavelengths. The above equations are valid in the range $G=3$--20~mag. Stars fainter than 20~mag are (routinely) observed but their end-of-mission number of transits resulting in usable data on ground may be reduced as a result of crowding, on-board priority management, on-board magnitude-estimation errors, finite on-board detection and confirmation probabilities, etc. Their standard errors will hence be (much) larger than predicted through this model. Stars brighter than $G = 3$~mag will be observed with a special mode with associated non-trivial calibration challenges \cite{2014SPIE.9143E..0YM}; the end-of-life astrometric standard errors for these stars will amount to at least several dozens of $\mu$as. For stars between $3$ and $12$~mag, reduced CCD integration times (through the use of Time-Delayed-Integration -- TDI -- gates) will be used to limit saturation. For these stars, the end-of-mission performance depends sensitively on the adopted TDI-gate scheme, which depends on CCD and on across-scan location inside the CCD, as well as on magnitude; in addition, on-board magnitude-estimation errors will cause a given bright star sometimes to be observed with the ``wrong'' TDI gate. The MAX function in Equation~(\ref{eq:z}) allows to ignore these ``complications'' and returns a constant parallax noise floor, at $\sigma_\varpi \sim 7~ \mu$as, for stars with $3 \leq G \leq 12.09$~mag.

The pre-launch relations between sky-average parallax standard error and position and proper-motion standard errors have not changed \cite{2012Ap&SS.341...31D}:
\begin{eqnarray}
\sigma_0   & = & 0.743 \cdot \sigma_\varpi;\\
\sigma_\mu & = & 0.526 \cdot \sigma_\varpi,
\end{eqnarray}
where $\sigma_0$ denotes the position error at mid-epoch in $\mu$as and $\sigma_\mu$ denotes the (annual) proper-motion error in $\mu$as~(yr$^{-1}$). The dependence of the astrometric errors on celestial coordinates as described in de Bruijne \cite*{2012Ap&SS.341...31D} is also still valid.

One should finally keep in mind that the current assessment exclusively takes the increased photon noise caused by the straylight into account. Since we do not have and hence cannot include a sophisticated calibration-error model, the performance estimates presented here do not include the (potential) additional performance degradation caused by increased estimation errors of the background levels around the stars, which are needed as input parameters in the image-parameter determination / centroiding / location estimation. Although the size of this additional degradation factor will not be known until significant stretches of real data have been processed and (iteratively) calibrated, it is expected that this effect will fit, even when combined with the transmission loss mentioned in Section~\ref{sec:com}, into the $20$\% ``science margin'' which has been included in all predictions.

\section*{Acknowledgements}

It is a pleasure to thank the conference organisers, in particular Nic Walton and Francesca Figueras, for the enjoyable meeting. Michael Davidson (Edinburgh) kindly provided the frequency distribution of the straylight levels measured during the commissioning of the Gaia spacecraft. The commissioning phase was supported by Gaia's prime contractor, Airbus Defence \& Space, and by Gaia's Mission Operations Centre (ESOC), Gaia's Science Operations Centre (ESAC), Gaia's Project Team (ESTEC), Gaia's Project Scientist Support Team (ESTEC), the Data Processing and Analysis Consortium (DPAC, in particular the Payload Experts), and the DPAC Project Office (PO).


\end{document}